\documentclass[aps,prx,twocolumn,tightenlines,superscriptaddress]{revtex4}

\usepackage{amssymb}
\usepackage{graphicx}
\usepackage{amsmath}
\usepackage{amsbsy}
\usepackage{amsthm}
\usepackage{bbm}
\usepackage{epsfig}
\usepackage{color}
\usepackage{hhline}

\newcommand{\beq}{\begin{eqnarray}}
\newcommand{\eeq}{\end{eqnarray}}

\begin{document}

\title{Witnessing Entanglement In Compton Scattering Processes Via Mutually Unbiased Bases}
\author{Beatrix C. Hiesmayr}
\email{Beatrix.Hiesmayr@univie.ac.at}
\affiliation{Faculty of Physics, University of Vienna, Boltzmanngasse 5, 1090 Vienna, Austria}
\author{Pawel Moskal}
\affiliation{Institute of Physics, Jagiellonian University, Cracow, Poland}

\begin{abstract}
We present a quantum information theoretic version of the Klein-Nishina formula. This formulation singles out the quantity, the \textit{a priori} visibility, that quantifies the ability to deduce the polarisation property of single photons. The Kraus-type structure allows a straightforward generalisation to the multiphoton cases, relevant in the decay of positronium which is utilized e.g. for metabolic {\bf PET}-imaging ({\bf P}ositron-{\bf E}mission-{\bf T}omograph). Predicted by theory but never experimentally proven, the two- or three-photon states should be entangled. We provide an experimentally feasible method to witness entanglement for these processes via {\bf MUB}s ({\bf M}utually {\bf U}nbiased {\bf B}ases), exploiting Bohr's complementarity. Last but not least we present explicit cases exemplifying the interrelation of geometry and entanglement including relations to its potentiality for teleportation schemes or Bell inequality violations or in future for detecting cancer in human beings.
\end{abstract}


\maketitle


\section{Introduction}
No doubt manifestations of entanglement are fascinating phenomena that have been witnessed for numerous physical systems at low and high energies. Moreover, new technologies based on entanglement are currently emerging. One such may be based on detecting cancer via the various types of entanglement manifesting in the two- or three-photon states of the decay process of positronium~\cite{PositroniumHiesmayr,Cancer1,Cancer2,Cancer3}, a bound state of an electron and its antiparticle. However, the theoretically predicted entanglement in those gammas has never been observed, because the energies are around the mass of an electron ($511keV$) and for such high energetic photons standard optical polarizers do not work. The new prototype J-PET ({\bf J}agelonian-{\bf P}ositron-{\bf E}mission {\bf T}omograph)~\cite{JPETT1,JPET1,JPETX,JPETT4,JPETT5,JPETT6} is based on plastic scintillators~\cite{scintillators} that shall be a key technology of a new generation of low cost and total-body scan PETs and, in addition, has shown in providing all key elements to detect the positronium~\cite{JPETT7,JPET3,JPETT8} and the Compton-scattered gammas~\cite{PolExp}. This paper shows how the entanglement can be witnessed and provides a concise quantum information theoretic framework for describing high energetic photons undergoing Compton scattering processes. If this step is taken, observables sensitive to entanglement may become visible in living beings along with all the well-known benefits of a standard PET-scan.

In detail the Klein-Nishina formula~\cite{KleinNishina} is reformulated in the open quantum formalism. Firstly, the scattering of a single photon in a scintillator is formulated in terms of an envelope function and a term in front of the polarisation interference, an \textit{a priory} visibility or interference contrast. This visibility is depending solely on the incoming energy $k_i$ of the photon and on the scattering angle $\tilde{\Theta}$ with the outgoing photon. We show that the visibility function is ruling the feasibility to measure polarisation effects via a Compton scattering process. It explains why above a certain energy no polarization properties can be attained and that for small angles and angles close to $180^\circ$ (backward scattering), independently of the energy, also no polarization effects are feasible. Furthermore, it tells the experimenter for which angles polarisation effects are measurable.

The open quantum formulation of the Klein-Nishina formula expressed in terms of Kraus-type operators~\cite{Kraus} allows a straightforward generalization to multiphoton states (separable or entangled states). This in turn also enables a straightforward adaption of the Kraus-type operators to derive the differential cross section for a given geometry. For instance, in the positronium decay into two or three photons the momentum vectors have to be back-to-back or in a plane, respectively.

Witnessing the entanglement in polarization  via the differential cross section is tricky, because also a separable state may lead to the same differential cross section, as we show in detail. In Section~\ref{Secwitnessingentnaglment} we therefore adapted a protocol~\cite{MUBHiesmayr} based on  MUBs ({\bf M}utually {\bf U}nbiased {\bf B}ases) that has been used to detect for the first time \textit{bound} entanglement of two photons entangled in their orbital momentum~\cite{MUBExpHiesmayr1,MUBExpHiesmayr2}.  Bound entanglement was predicted in 1998 by the Horodecki family~\cite{HorodeckiBound} and is a curious type of entanglement that cannot be distilled, i.e. no pure maximally entangled states can be gained via local unitary operations and classical communication (LOCC). Other recent experiments with photons exploiting MUBs can be found e.g. in Refs.~\cite{MUBexp1,MUBexp2,MUBexp3}.

Observables that are mutually unbiased cannot show maximal correlation in all basis choices except for the states that are entangled, this property, also often phrased as Bohr's complementary, is exploited to detect entanglement. In addition, the information theoretic formulation allows in interpreting the joint scattering cross sections in terms of probabilities where the visibility is the ruling quantity. This analogy allows to connect the cross section with typical quantum protocols such as teleportation~\cite{teleportation,teleportation2} or violations of a Bell inequality~\cite{Bell,chsh}.

The paper is organized as follows. In Section~\ref{Secsinglephoton} we provide a brief introduction to the notorious difficult single photon description. In the next Section~\ref{SecKraus} we present the Kraus representation of the Klein-Nishina formula providing us a unified framework to address the quantum information theoretic questions. The application to more than one photon undergoing a Compton scattering process is given in Section~\ref{Secapplication}. This shows the need of developing tools to distinguish between separable and entangled states provided in Section~\ref{Secwitnessingentnaglment}. In Section~\ref{secfeasibility} we connect the MUB-witness to the experimental setup. Last but not least we present a case with non-trivial geometry, e.g. the decay of positronium into three photons in Section~\ref{Secnontrivial} followed by a conclusion and outlook. Furthermore the paper is equipped with two appendices covering the details on a two-photon wave function and symmetry consideration as well as another witness suitable to detect entanglement in the positronium decay.

\begin{figure*}
\includegraphics[width=0.8\textwidth]{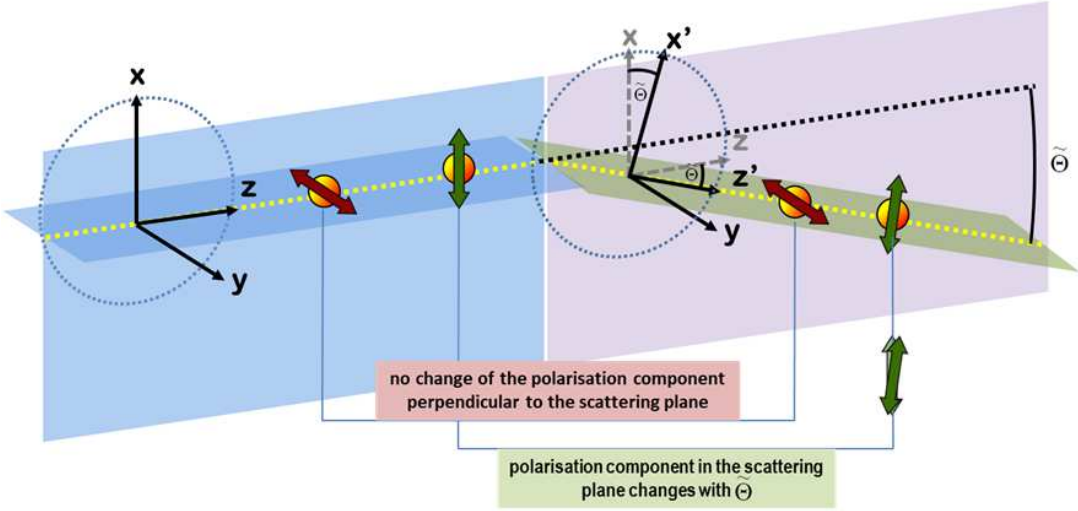}
\caption{(Color online) This graphic shows the Compton scattering process for incoming linear polarized photons with respect to the scattering plane $\phi=\phi'$. The component perpendicular to the scattering plane does not undergo a change in the Compton process. The polarisation component in the scattering plane has to change since the direction of motion changes and the polarisation components have to be perpendicular to the direction of motion. In our particular reference frame by defining $H$ to be the polarisation component in the scattering plane and $V$ being the component perpendicular to the scattering plane, we find from formula~(\ref{generalscatteringangles}) the following transition amplitudes $H\longrightarrow H': \cos\tilde{\Theta},\; V\longrightarrow V': -1,\; H\longrightarrow V': 0,\; V\longrightarrow H': 0$.}
	\label{visualizationpolarisationscattering}
\end{figure*}

\section{Single photon description}\label{Secsinglephoton}

Photons are spin-$1$, massless and relativistic particles. We describe a photon by its energy $E_\gamma=\hbar \omega= c \hbar |\bf{k}|$, its direction of propagation $\hat{\bf{k}}=\frac{\bf{k}}{|\bf{k}|}$ and its polarisation state $|\lambda\rangle$. Those three properties are known to characterize the quantum particle ``photon'' fully, if in addition the transversality condition holds, ${\bf \hat{k}}\cdot{\boldsymbol \varepsilon}({\bf \hat{k}},\lambda)=0$, where the (complex) three dimensional polarisation vector $\boldsymbol{\varepsilon}(\bf{\hat{k}},\lambda)\equiv|\lambda\rangle$ fulfills the Poincar\`{e} transformations, relating the spin degrees of freedom with the rotations in our real three-dimensional space $\mathcal{R}^3$. For concise mathematical frameworks capturing the quantum information content of single photons see e.g.~Refs.\cite{SinglePhoton,HiesmayrSinglePhoton}. For a photon propagating in direction (spherical coordinates)
\beq
\bf{\hat{k}}&=&\frac{\bf{k}}{|\bf{k}|}\;=\; \left(\begin{array}{c} \sin\theta \cos\phi\\ \sin\theta \sin\phi\\ \cos\theta\end{array}\right)\;,
\eeq
the corresponding three-dimensional complex polarisation vector
\beq
\boldsymbol{\varepsilon}(\bf{\hat{k}},\lambda)&=&\frac{1}{\sqrt{2}} \left(\begin{array}{c} -\lambda \cos\theta \cos\phi+i \sin\phi\\ -\lambda\cos\theta \sin\phi-i \cos\phi\\ \lambda\sin\theta\end{array}\right)
\eeq
can be defined, which characterises the two circular polarised eigenstates, $|R\rangle\equiv {\boldsymbol\varepsilon}({\bf\hat{k}},\lambda=+1)$ and $|L\rangle\equiv {\boldsymbol\varepsilon}({\bf \hat{k}},\lambda=-1)$, via the eigenvalue equation $i \hat{\bf{k}}\times\boldsymbol{\varepsilon}(\bf{\hat{k}},\lambda)=\lambda\; \boldsymbol{\varepsilon}(\bf{\hat{k}},\lambda)$.
The linear polarised states (with respect to the propagation direction) can be defined by equal superposition of the two circular polarized states $\lambda=\pm1$
\begin{widetext}
\beq
\boldsymbol{\varepsilon}_H(\hat{\bf{k}})&=&\frac{e^{i\xi}}{\sqrt{2}}\left\{\boldsymbol{\varepsilon}(\hat{\bf{k}},+1)-\boldsymbol{\varepsilon}(\hat{\bf{k}},-1)\right\}
\;=\;(-e^{i\xi})\left(\begin{array}{c}\cos\theta\cos\phi\\ \cos\theta \sin\phi\\-\sin\theta\end{array}\right)\;\Longleftrightarrow\;|H\rangle\;=\;\frac{1}{\sqrt{2}}\{|R\rangle-|L\rangle\}\nonumber\\
\boldsymbol{\varepsilon}_V(\hat{\bf{k}})&=&\frac{e^{i\xi}}{\sqrt{2}}\left\{\boldsymbol{\varepsilon}(\hat{\bf{k}},+1)+\boldsymbol{\varepsilon}(\hat{\bf{k}},-1)\right\}\;=\;(-i e^{i\xi})\left(\begin{array}{c}-\sin\phi\\ \cos\phi\\0\end{array}\right)\;\Longleftrightarrow\;i\,|V\rangle\;=\;\frac{1}{\sqrt{2}}\{|R\rangle+|L\rangle\}\;.
\eeq
\end{widetext}
Let us remark here that for any direction of the photon one can always define one of the two orthogonal linear polarised states to be independent of the polar angle $\theta$. Obviously, choosing the $z$-axis as the reference frame of propagation($\theta=0$) also the second linear polarised states has a zero $z$-component and, consequently, the linear polarised states are confined to the $x,y$-plane. For a single photon one may always choose such a reference system, however, as we show hereafter, this fails if more photons are considered.  Without loss of generality one can fix the overall phase $\xi$ to e.g. $\xi=0$.

From the classical point of view the $x$-part and $y$-part of the polarisation vector $\boldsymbol{\varepsilon}$ can be defined to represent the electric field of the electromagnetic wave and thus the polarisation property of photons (a propagation in $z$-direction is assumed). From the quantum mechanical point of view one can identify the oscillation of the electric field vector in $x$-direction with the horizontal polarised state $|H\rangle$ and the $y$-direction with the vertical polarised state $|V\rangle$.

\section{Kraus representation of the Klein-Nishina formula}\label{SecKraus}

The well-known Klein-Nishina formula~\cite{KleinNishina} relates an incoming polarized photon, described by its direction ${\bf\hat{k}}_i$, its energy $k_i$ [choosing proper units $[k]=[\frac{m_e c^2}{c\hbar}]\equiv 1$; corresponding to $511$keV photons] and its polarisation vector $\boldsymbol{\varepsilon}$, to an outgoing polarized photon described by its direction $\bf{\hat{k}}'$, its energy (that depends on $k_i$ and the scattering angle $\tilde{\Theta}$)  and its polarisation vector $\boldsymbol{\varepsilon}'$ is given by
\beq
\sigma_{jl}&=&\frac{r_0}{2}\left(\frac{k'(k_i,\tilde{\Theta})}{k_i}\right)^2\cdot\left(\gamma(k_i,\tilde{\Theta})-2+4 |{\boldsymbol \varepsilon}_j'^*\cdot{\boldsymbol \varepsilon}_l|^2\right)\nonumber\\
\eeq
where $l/j$ refer to the initial/final polarised states and  $r_0$ is classical electron radius. The energy-depending part is given by
\beq
&&\gamma(k_i,\tilde{\Theta})\;=\;\frac{k_i}{k'(k_i,\tilde{\Theta})}+\frac{k'(k_i,\tilde{\Theta})}{k_i}\nonumber\\
&&\textrm{with}\quad k'(k_i,\tilde{\Theta})\;=\;\frac{1}{1-\cos\tilde{\Theta}+\frac{1}{k_i}}\;,
\eeq and the scattering angle $\tilde\Theta=\tilde\Theta(\theta,\phi)$ by
\beq
\cos\tilde{\Theta}&:=&\hat{{\bf k}_i}\cdot\hat{\bf{k'}}\;=\;\cos\theta'\cos\theta+\cos(\phi-\phi')\sin\theta'\sin\theta\;.\nonumber
\eeq
A classical and quantum mechanical discussion of the polarisation dependent part can be found in Ref.~\cite{Fanno}. Let us write down explicitly the four different amplitudes with respect to the linear polarised basis
\beq\label{generalscatteringangles}
H&\rightarrow& H':f_{HH}=\boldsymbol{\varepsilon}_H'^*\cdot\boldsymbol{\varepsilon}_H\nonumber\\
&&=\;\cos\theta'\cos\theta\cos(\phi-\phi')+\sin\theta'\sin\theta\nonumber\\
&&=\;\sqrt{\cos^2\tilde{\Theta}-\frac{1}{2}(\cos(2\theta)+\cos(2\theta'))\sin(\phi-\phi')^2}\nonumber\\
H&\rightarrow& V':f_{VH}=\boldsymbol{\varepsilon}_V'^*\cdot\boldsymbol{\varepsilon}_H=i\cos\theta \sin(\phi-\phi')\nonumber\\
V&\rightarrow& H':f_{HV}=\boldsymbol{\varepsilon}_H'^*\cdot\boldsymbol{\varepsilon}_V=-i \cos\theta' \sin(\phi-\phi')\nonumber\\
V&\rightarrow& V':f_{VV}=\boldsymbol{\varepsilon}_V'^*\cdot\boldsymbol{\varepsilon}_V=-\cos(\phi-\phi')\;,
\eeq
for which we observe that in the case one chooses the scattering plane, i.e $\phi=\phi'$, one is left with only two non-zero contributions. This  gives raise to the intuitive picture of the Compton scattering process visualized in Fig.~\ref{visualizationpolarisationscattering}. It shows that only the polarisation component in the scattering plane changes (by $\cos\tilde\Theta=\cos(\theta-\theta')$) due to change in the propagation direction.

Throughout the paper we will assume that the Compton interaction in the material is not sensitive to the outgoing polarisation, as it is the case for most materials such as the plastic scintillators of J-PET~\cite{scintillators}. Therefore, we have to sum over all final possibilities.
Furthermore, we rewrite the Klein-Nishina formula in a Kraus operator representation, i.e.
\beq
\sigma_\rho
&=& \frac{r_0}{2} \left(\frac{k'(k_i,\tilde{\Theta})}{k_i}\right)^2 \left\lbrace Tr(\mathcal{K}_1\;\rho\;\mathcal{K}_1^\dagger)+ Tr(\mathcal{K}_2\;\rho\;\mathcal{K}_2^\dagger)\right\rbrace\nonumber\\
\eeq
with $\rho$ being the initial (pure or mixed) state and the two Kraus-type operators defined  are given by (with respect to the linear polarised basis $\{H,V\}$)
\beq
\mathcal{K}_{1}&=&\sqrt{\gamma(k_i,\tilde{\Theta})-2}\;\; \mathbbm{1}_2\\
\mathcal{K}_{2}&=&\sqrt{2} \left(\begin{array}{c c} f_{HH}&f_{HV}\\
f_{VH}&f_{VV}\end{array}\right)
\eeq
Note that the completeness relation $\sum_i \mathcal{K}_{i}^\dagger \mathcal{K}_{i}= \mathbbm{1}$ does not hold since we factored out the common ratio of the incoming and outgoing photon's energy.

\begin{figure*}
(a)\includegraphics[width=0.47\textwidth]{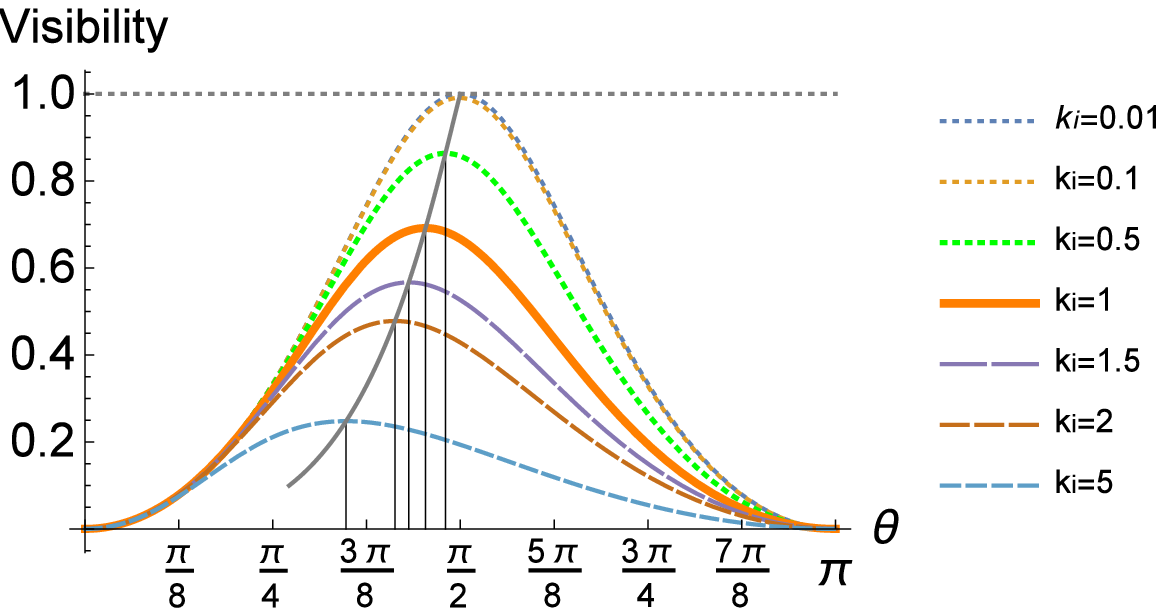}
(b)\includegraphics[width=0.47\textwidth]{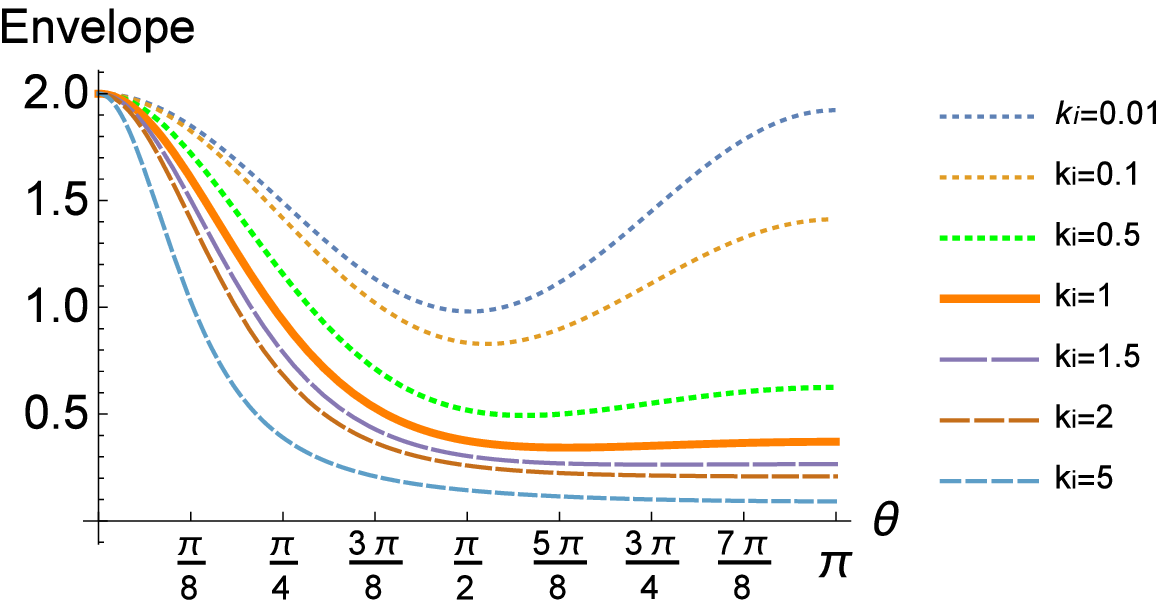}
\caption{(Color online) The plots in (a) show the visibility $\mathcal{V}(\tilde{\Theta},k_i)$ in dependence of Compton scattering angle $\tilde{\Theta}$ for different initial energies $k_i$. For a $k_i=1\equiv 511keV$, e.g. photons emerging from a positronium decay, the maximum value of the visibility reaches its maximum of $0.69$ at $\tilde{\Theta}=81,67^\circ$. The vertical lines mark the angles for which the visibility maximizes. The plots in (b) show the envelope function $\mathcal{F}(\tilde{\Theta},k_i)$ in dependence of  $\tilde{\Theta}$ and $k_i$. Thus there is a tradeoff between statistic (high value of $\mathcal{F}(\tilde{\Theta},k_i)$) and the maximum of the visibility $\mathcal{V}(\tilde{\Theta},k_i)$.}
	\label{figvisitbilityenergy}
\end{figure*}

Via this open quantum formulation we obtain an information theoretic form for any pure or mixed state $\rho=\sum_{i,j=H}^V\rho_{ij}\,|i\rangle\langle j|$ of the differential cross section
\begin{widetext}
\beq
\lefteqn{\sigma_\rho(\tilde{\Theta},\Phi,k_i)\;=}\nonumber\\
&&r_0\; \mathcal{F}(\tilde{\Theta},k_i)\cdot\overbrace{\frac{1}{2}\left\lbrace 1-\mathcal{V}(\tilde{\Theta},k_i)\cdot \left(\underbrace{(\rho_{HH}-\rho_{VV})}_{\textrm{\textcolor{green}{linear pol. part}}}\;\cos(2\Phi)+\underbrace{2\,Im\{\rho_{HV}\}}_{\textrm{\textcolor{green}{circular pol. part}}}\;\sin(2\Phi)\right)\right\rbrace}^{\textit{\large{\textcolor{red}{\textsl{probability part}}}}}\;,
\eeq
\end{widetext}
where $\Phi$ being the relative angle between the plane spanned by the basis states $\{|H\rangle,|V\rangle\}$ and the scattering plane.
Here the two interference terms characterize the linear and the circular polarisation part, respectively. For the third mutually unbiased bases, $\{|45^\circ\rangle,|-45^\circ\rangle\}$ ($\rho_{HH}=\rho_{VV}=\frac{1}{2}$ and $\rho_{HV}=\rho_{VH}=\pm\frac{1}{2}$), the interference term vanishes, i.e. the differential cross section equals the \textit{shaping} or \textit{envelope} function $\mathcal{F}(\tilde{\Theta},k_i)$ of the scattering process
\beq
\mathcal{F}(\tilde{\Theta},k_i)&:=&\left(\frac{k'(\tilde{\Theta},k_i)}{k_i}\right)^2\cdot   (\gamma(k_i,\tilde{\Theta})-\sin^2\tilde{\Theta})\;.
\eeq
The term $\mathcal{V}$ in front of the interference terms is the \textit{a priory} visibility of the scattering process
\beq
\mathcal{V}(\tilde{\Theta},k_i):=\frac{\sin^2\tilde{\Theta}}{\gamma(k_i,\tilde{\Theta})-\sin^2\tilde{\Theta}}
\;,
\eeq
it defines a kind of interference contrast. Since it multiplies the interference term, only a considerable non-zero value of the visibility allows in deducing the polarisation via the Compton scattering process. In Fig.~\ref{figvisitbilityenergy} the visibility and the envelope function are plotted in dependence of the scattering angle $\tilde{\Theta}$ for different initial photon energies $k_i$. We observe that, independently of the energy, if the photon is scattered under small angles or close to $180^\circ$, the visibility is close to zero, i.e. the scattering process does not become sensitive to the polarisation state of the incoming photon. A similar behaviour is found for the Mott scattering, i.e. Rutherford scattering with identical particles. In general, as shown in Ref.~\cite{ComplementarityHiesmayr}, double slit-like experiments attain a unified description in terms of the \textit{a priori} visibility and the \textit{a priori} predictability~\cite{Yasin,Englert,Jaeger,Vaccaro}, the ``\textit{which way}'' information. This includes the decay of neutral mesons, spin systems in magnetic fields or typical interferometric setups or the decay of hyperon~\cite{HyperonHiesmayr}. The \textit{a priori} visibility may be interpret to present the ``wave particle'' property, whereas the predictability $\mathcal{P}$ captures the ``particle property'', and Bohr's complementarity relation becomes quantified since the general relation $\mathcal{V}^2+\mathcal{P}^2\leq 1$ has to hold (equality holds for pure states).

The envelope function (with proper normalization) shows that the scattering under small angles is more probable than for greater angles and a small increase is found for angles close to $\pi$. The maximum of the visibility is strongly energy dependent and diverges for increasing energies from $90^\circ$. For instance, for an incoming $k_i=1\equiv 511keV$ photon the maximal interference contrast is $\mathcal{V}(81,67^\circ,1)=0.69$, i.e. in a scan over the azimuthal angle $\Phi$ the maxima and minima differ by 69\%.

In summary, the higher the energy the less accurate is the determination of the initial polarisation via Compton scattering and the optimal scattering angles diverge from $90^\circ$. In a next step we discuss the multi-photon case and, in particular, the two-photon case in detail.

\section{Application of the Kraus operator formalism to the case of two incoming photons}\label{Secapplication}

The advantage of the Kraus-representation for the Compton scattering is that it allows a straightforward generalization, i.e. for a system of $z$ photons in any given state $\rho$ the cross section is derived to
\beq\label{generalmultipartiteresult}
\sigma_{\rho}&=&(\frac{r_0}{2})^z \left(\frac{k_a}{k_{i_a}}\right)^2 \left(\frac{k_b}{k_{i_b}}\right)^2\cdots \left(\frac{k_z}{k_{i_z}}\right)^2 \sum_{l_a,l_b,\dots l_z=1}^2\nonumber\\
\lefteqn{Tr(\mathcal{K}_{l_a}^{(a)}\otimes \mathcal{K}_{l_b}^{(b)}\cdots \otimes \mathcal{K}_{l_z}^{(z)}\;\rho\;\mathcal{K}_{l_a}^{(a)\dagger}\otimes \mathcal{K}_{l_b}^{(b)\dagger}\cdots \otimes \mathcal{K}_{l_z}^{(z)^\dagger})}\nonumber\\
\eeq

Let us discuss here an explicit example, the decay of para-positronium into two photons. Momentum conservation restricts the two photons to be back-to-back in the rest systems of the para-positronium and energy conservation implies that both photons need to have equal energies, namely $511keV$ corresponding to  $k_i=k_{i'}=1$, the mass of an electron or a positron. Taking advantage of the open quantum formalism, we apply this geometrical restriction to the Kraus operators, i.e. the direction of the incoming photon $a$ is described by the polar coordinates $(\theta,\phi)$ with respect to some fixed coordinate system and the scattered photon by $(\theta_a,\phi_a)$, then the polar angles describing the direction of the second incoming photon $b$ have to be $(\pi-\theta,\phi+\pi)$, whereas its scattered photon is described by $(\theta_b,\phi_b)$. Note that this back-to-back kinematic connects the two involved Kraus operators $\mathcal{K}^{a/b}$, it singles out a particular direction (the one of the propagation), but the linear polarised states in the plane perpendicular to this direction of motion can still be defined arbitrarily. Of course, a phase match occurs in the case a joint initial state is considered.

\begin{widetext}
Applying the general formula~(\ref{generalmultipartiteresult}) for two back-to-back photons and choosing one of the four mutually orthogonal maximally entangled Bell states, i.e. $|\psi^\pm\rangle_{\textrm{lin}}=\frac{1}{\sqrt{2}}\{|HV\rangle\pm|VH\rangle\},|\phi^\pm\rangle_{\textrm{lin}}=\frac{1}{\sqrt{2}}\{|HH\rangle\pm|VV\rangle\}$, we obtain the result (choice $\theta=0$, propagation in $z$-direction, $\phi$ arbitrary)
\beq\label{crosssectionstwophotons}
\sigma_{|\psi^\alpha\rangle_{\textrm{lin}}/|\phi^\alpha\rangle_{\textrm{lin}}}&=& r_0^2\; \mathcal{F}(k_i,\tilde{\Theta}_a) \mathcal{F}(k_{i},\tilde{\Theta}_b)\cdot\frac{1}{4}\biggl\lbrace 1 \mp\mathcal{V}(k_i,\tilde{\Theta}_a)\cdot \mathcal{V}(k_{i},\tilde{\Theta}_b)\cdot \cos(2((\phi_a-\phi)\mp\alpha(\phi_b-\phi)))\biggr\rbrace\;.
\eeq
The visibilities of both scattering processes multiply, which is a consequence of the tensor product structure of the Kraus operators. Obviously, for any multi-photon process the total visibility always decreases with the power of the single visibilities. Similar observations hold for the envelope functions.

Only for the states, $|\psi^+\rangle_{\textrm{lin}},|\phi^-\rangle_{\textrm{lin}}$  the scattering cross sections of Eq.~(\ref{crosssectionstwophotons}) is independent of the choice $\phi$, the assigned $x$- and $y$-axes to the two-particle system. So far, we have neglected the Bose symmetry. As the detailed computations in the Appendix~\ref{appendix1} show these two Bell states are exactly those obeying the Bose symmetry. If we are interested in the decay of para-positronium with total angular momentum $L=0$ we have a further symmetry that is conserved, namely parity. In its ground state para-positronium has partity $\mathcal{P}=-1$ which is only obtained for $|\psi^+\rangle_{\textrm{lin}}=|\phi^-\rangle_{\textrm{circ}}$.

In summary the two photons resulting from the decay process of para-positronium  are ruled by three physical invariant quantities, the two scattering angles $\cos\tilde{\Theta}_{a}={\bf k}_{i}\cdot{\bf k}_{a}, \cos\tilde{\Theta}_{b}={\bf k}_{i'}\cdot{\bf k}_{b}$ and the angle between the two scattering planes $\eta=\phi_a-\phi_b$. In particular the choice of the phase $\phi$ should not be experimentally observable. \textit{Is this, however, enough to experimentally prove that entanglement is required to describe the experimental observed differential cross section?}

For that reasons let us assume that the initial two-photon state is separable. For a full basis set of separable states, e.g. $\{|HH\rangle,|HV\rangle,|VH\rangle,|VV\rangle\}$, we find from Eq.~(\ref{generalmultipartiteresult}) utilizing the same Kraus operators
\beq\label{crosssectionSEP}
\sigma_{|HH\rangle/|VV\rangle}&=& r_0^2\; \mathcal{F}(k_i,\tilde{\Theta}_a) \mathcal{F}(k_{i},\tilde{\Theta}_b)\cdot\frac{1}{4}(1\mp\mathcal{V}(k_i,\tilde{\Theta}_a)\cos(2(\phi_a-\phi)))\cdot(1\mp\mathcal{V}(k_i,\tilde{\Theta}_b)\cos(2(\phi_b-\phi)))\\
\sigma_{|HV\rangle/|VH\rangle}&=& r_0^2\; \mathcal{F}(k_i,\tilde{\Theta}_a) \mathcal{F}(k_{i},\tilde{\Theta}_b)\cdot\frac{1}{4}(1\mp\mathcal{V}(k_i,\tilde{\Theta}_a)\cos(2(\phi_a-\phi)))\cdot(1\pm\mathcal{V}(k_i,\tilde{\Theta}_b)\cos(2(\phi_b-\phi)))
\eeq
\end{widetext}
Let us assume that the source produces, e.g. a mixed state $\rho=\frac{1}{2}|HV\rangle\langle HV|+\frac{1}{2}|VH\rangle\langle VH|$ obeying the Bose symmetry and choosing $\phi=\phi_b$, then, obviously, we obtain the same result as for an initial maximal entangled state $|\psi^+\rangle_{\textrm{lin}}$, i.e. Eq.~(\ref{crosssectionstwophotons}). Thus, without invoking the parity conservation (which is only the case for the specific decay of para-positronium into two photons in its ground state) both cases are not distinguished by measuring the cross section $\sigma$.  \textit{How should one distinguish experimentally these two distinct possibilities from each other?} This we discuss in the following sections, by introducing firstly an entanglement witness based on mutually unbiased  bases and then applying it to the setting in J-PET experiment.

\section{Witnessing Entanglement}\label{Secwitnessingentnaglment}

 \begin{figure*}
\begin{center}(a)\includegraphics[width=0.47\textwidth]{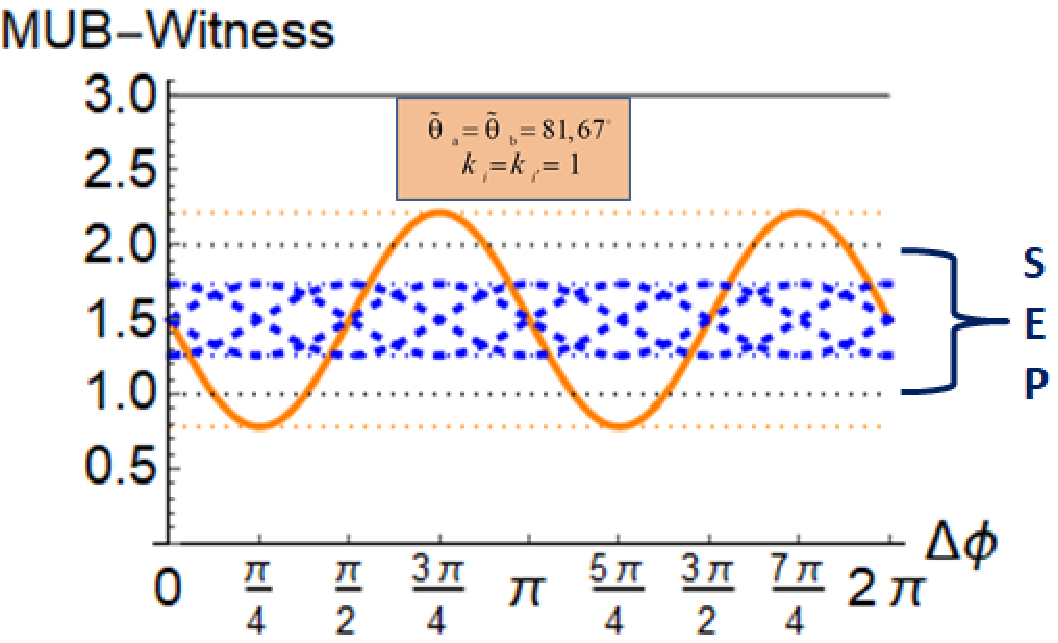}
(b)\includegraphics[width=0.47\textwidth]{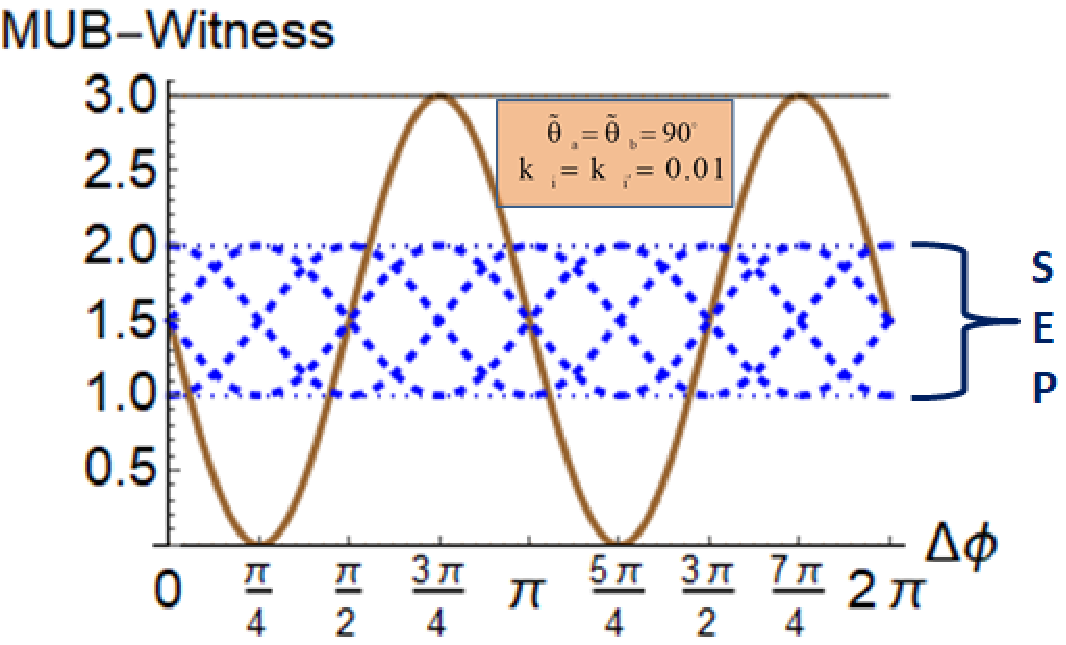}
\end{center}
\caption{(Color online) The plots show the MUB-witness $I_3$ for highest possible visibilities for different initial states for (a) two $k_i=k_i'=1\equiv 511keV$ photons and for (b)  $k_i=k_i'=0.01\equiv 5110eV$ photons in dependence of $\Delta\phi=\phi_a-\phi_b$ (and a fixed reference coordinate system $\theta,\phi$). The horizontal solid black lines show the optima for the MUB-witness $I_3$ for entangled states, Eq.~(\ref{IsymENT}), and the dotted black lines those for separable states, Eq.~(\ref{IsymSEP}). The dashed lines present the optima for the reduced visbililities, respectively. The solid (a) orange and (b) brown curve corresponds to the result for the initial maximally entangled state $|\psi^+\rangle$. The dashed blued curves present the result for $I_3$ for different separable states undergoing the Compton scattering process.}
	\label{MUBwitness}
\end{figure*}

Entanglement manifests itself in correlations that are stronger than correlations that can be obtained by classical or separable systems. The crucial point is that the correlation may be found maximal for a particular setup, but only in the case maximal correlations are found also for other setups without adaptation, then one outperforms classical systems. This happens particularly for mutually unbiased setups. Let us label a complete set of orthonormal basis vectors by $\{|i_l^{(k)}\rangle\}_{l=0}^{d-1}$ with $d$ being the dimension of the system, i.e. the number of possible outcomes. Another complete set of other basis vectors $\{|i_l^{(k')}\rangle\}_{l=0}^{d-1}$ is called mutually unbiased if and only if the relation $|\langle i_l^{(k)}| i_{l'}^{(k')}\rangle|^2 \;=\; \delta_{l,l'}\delta_{k,k'}+(1-\delta_{k,k'})\frac{1}{d}$ holds.

Let us consider two parties, Alice and Bob, that each observe the azimuthal distribution $\phi_a/\phi_b$ of one photon for fixed polar scattering angles $\tilde{\Theta}_a/\tilde{\Theta}_b$. Actually, in our case one gathers all events and sorts them accordingly since there is no active control over the scattering angles for each event. In order to relate their results Alice and Bob need to agree on a particular coordinate system ($\theta,\phi$). Comparing their outcomes pairwise they obtain a distribution depending on $\phi_a,\phi_b,\theta,\phi$ (for fixed  $\tilde{\Theta}_a,\tilde{\Theta}_b$).

Let us now construct a function that quantifies the correlation, namely a \textit{fully correlated} system should be one if knowing the outcome in one subsystem one can predict with certainty the outcome of the other subsystem. On the other hand a \textit{fully uncorrelated} system should be one knowing the outcome in one subsystem does not tell us anything about the outcome of the other system, it is totally random. Such a function is for instance the sum of two joint probabilities $\mathcal{C}_{i,j}:=P(i_1,j_1)+P(i_2,j_2)$ for the chosen measurement basis $\{|i_1\rangle,|i_2\rangle\}$ for Alice and $\{|j_1\rangle,|j_2\rangle\}$ for Bob. Obviously, for a fully correlated system, classical or quantum, $\mathcal{C}_{i,j}=1$ holds. On the other hand for a fully uncorrelated system to obtain the result $i_1$ is $\frac{1}{2}$ and to obtain the result  $j_1$ is independent of $i_1$, hence also $\frac{1}{2}$. The same holds true for the joint results $i_2,j_2$, thus the correlation function gives $\mathcal{C}_{i,j}=\frac{1}{4}+\frac{1}{4}=\frac{1}{2}$. Thus the function $\mathcal{C}_{i,j}\in[\frac{1}{2},1]$ quantifies the probability of predicting the measurement outcomes of the dichotomic measurement $i$ if the outcome of the dichotomic measurement $j$ is known and vice versa. Let us remark here that an optimization of labelling of the outcomes is assumed (else the function can be as well zero) and it works for any dimension $d$. Furthermore, let us point out that this function is per se independent of considering classical or quantum systems.

For example, if we have a fully entangled state, e.g. $|\psi^+\rangle_{\textrm{lin}}$, the joint probability is computed by the standard quantum rules. We obtain the maximum of the correlation function $\mathcal{C}$ if we choose for Alice and Bob the linear polarized bases, i.e. $i=j=\{H,V\}$ and $i_1=j_2=H; i_2=j_1=V$ (thus an optimal correlation function needs an optimisation over the unphysical labels).

A separable state as e.g. $|HV\rangle,|VH\rangle$ or even $\frac{1}{2}|HV\rangle\langle HV|+\frac{1}{2}|VH\rangle\langle VH|$, however, results also in $\mathcal{C}=1$, since these systems are also maximally correlated. Thus this correlation function does not tell us something concerning separability versus entanglement in the case of a single measurement (corresponding to one mutually unbiased basis). Yet, as generally shown for any bipartite $d$-dimensional system, indeed it can be exploited to detect entanglement versus separability~\cite{MUBHiesmayr}. The point is that for another measurement choice, being mutually unbiased to the first choice,  the correlation function for a maximal entangled state can obtain again the maximum value, whereas for a separable state only the value $\frac{1}{d}$ is possible. Correlations for separable system may be chosen maximal for one measurement setup but this limits the possibilities for any other MUB setup. This means adding the results from different MUB setups gives in general lower and upper bounds, which may be violated for an entangled state. Generally, witnesses based on MUBs have turned out to be powerful in detecting entanglement~\cite{MUBExpHiesmayr1,MUBExpHiesmayr2,MUBDariusz} and quite same considerations hold true if one uses symmetric informationally complete positive operator-valued measures (SICs), see Ref.~\cite{BaeSIC}, because in the case of complete sets MUBs or SICs are related by a quantum $2$-design~\cite{kdesignHiesmayr}.

We focus now on MUBs, but similar considerations hold true for the SICs which are given in detail in the Appendix~\ref{appendixSIC}. For our two-level systems and choosing $m$ mutually unbiased bases the inequality~\cite{MUBHiesmayr,kdesignHiesmayr}
\beq
\frac{m-1}{2}\leq I_m(\rho)\leq 1+\frac{m-1}{2}
\eeq
holds for any separable state, where
\beq\label{MUBwitnessFormula}
I_m(\rho)&=&\sum_{k=1}^{m} C_{i^{(k)} j^{(k)}}(\rho)=\sum_{k=1}^{m}\sum_{l=1}^{2}\;P(i_l^{(k)},j_l^{(k)})\nonumber\\
&\stackrel{QM}{=}&\sum_{k=1}^{m}\sum_{l=1}^{2}\;Tr\left\{|i_l^{(k)} j_l^{(k)}\rangle\langle i_l^{(k)} j_l^{(k)}|\;\rho\right\}\;.\eeq
The index $k$ enumerates the MUB setups of Alice and Bob. In case of two-dimensional systems the maximal number of a complete set of MUBs is known to be three, e.g. the eigenvectors of the three Pauli matrices. For high dimensional systems this changes drastically.

In the case of the Compton scattering we are summing over all final states, thus we cannot choose different setups, however, we can reverse the argumentation by varying instead the initial states. Suppose the source produces a maximally entangled state  $|\psi^+\rangle_{\textrm{lin}}$, then when Alice and Bob plot their joint results in dependence of $\Delta \phi=(\phi_a-\phi)-(\phi_b-\phi)=\phi_a-\phi_b$ (which should correspond to the theoretical result~(\ref{crosssectionstwophotons})). They interpret one of the two maxima belonging to the case that Alice's particle has ``$H$'' and Bob's particle ``$V$'' with respect to the fixed phase $\phi$, whereas the other maximum is the vice-versa case. On the other hand, they can as well claim that the source produces the state  $|\phi^-\rangle_{\textrm{circ}}$ since $|\psi^+\rangle_{\textrm{lin}}=|\phi^-\rangle_{\textrm{circ}}$. Then one of the two maxima corresponds to ``$RR$'' and the other one to  ``$LL$''. The same reasoning works for the state $|\phi^-\rangle=\frac{1}{\sqrt{2}}\{|{\tiny +}45^\circ{\tiny +}45^\circ\rangle-|{\tiny -}45^\circ{\tiny -}45^\circ\rangle\}$. All these three states represent the same physical state, i.e. it has been only rewritten in different mutually unbiased basis choices. Consequently, all three correlation functions give the maximum value of $1$, thus $I_3=1+1+1=3$. Differently stated, the obtained joint differential cross section can be interpreted in three different ways.

On the other hand if one considers a separable state, say $|HV\rangle$, then the correlation function depends on the fixed phases $\phi$, see Eq.~(\ref{crosssectionSEP}), but may be chosen to give the maximal value. If one chooses another separable state, e.g. $|RL\rangle$, then again the correlation function can be chosen to give the maximal value, but for another fixed $\phi$. Now summing up three mutually unbiased setups of separable states and optimizing over the choice of $\phi$ we find $I_3(\rho_{\textrm{SEP}})\in[1,2]$. This is in strong contrast to entangled states which generally lead to $I_3(\rho_{\textrm{ENT}})\in[0,3]$. For these considerations we have used the maximal visibility of $1$. Consequently, in the case of rotation symmetry invariant states for a general setup with an \textit{a priory} visibility we have
\beq\label{IsymENT}
I_3(\rho_{\textrm{ENT}})&\in&\left[\frac{1}{2}(3-3\mathcal{V}(k_i,\tilde{\Theta}_a)\mathcal{V}(k_i',\tilde{\Theta}_b)),\right.\nonumber\\
&&\left.\qquad\frac{1}{2}(3+3\mathcal{V}(k_i,\tilde{\Theta}_a)\mathcal{V}(k_i',\tilde{\Theta}_b))\right]\\
\label{IsymSEP}
I_3(\rho_{\textrm{SEP}})&\in&\left[\frac{1}{2}(3-\,\mathcal{V}(k_i,\tilde{\Theta}_a)\mathcal{V}(k_i',\tilde{\Theta}_b)),\right.\nonumber\\
&&\left.\qquad\frac{1}{2}(3+\,\mathcal{V}(k_i,\tilde{\Theta}_a)\mathcal{V}(k_i',\tilde{\Theta}_b))\right]
\eeq
for the normalized part. We observe that if $\mathcal{V}(k_i,\tilde{\Theta}_a)\mathcal{V}(k_i',\tilde{\Theta}_b)$ is greater than $\frac{1}{3}$, the entanglement is witnessed directly by $I_3$.
Here the assumption was added that separable states undergo the same damping dynamics due to the Compton scattering process as entangled states, this means that any value greater than the optimum of $I_3(\rho_{\textrm{SEP}})$ witnesses entanglement of a two-photon state. This is visualized for low and high energies of the bipartite photonic system in Fig.~\ref{MUBwitness}.

\begin{figure}
\includegraphics[width=0.5\textwidth]{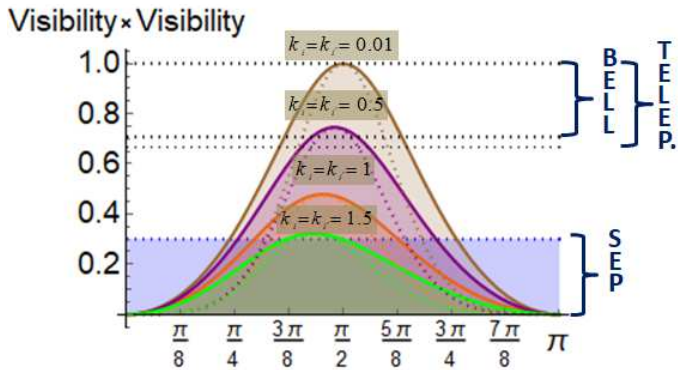}
\caption{(Color online) These plots show the functions $V(k_i,\tilde\Theta_a)\cdot V(k_i,\tilde\Theta_b)$ in dependence of $\tilde\Theta_a$ for different energies of the two photons. The solid curves correspond to $V(k_i,\tilde\Theta_a)\cdot\textrm{max}\{V(k_i,\tilde\Theta_b)\}$, whereas the dashed curves represent the case of $\tilde\Theta_a=\tilde\Theta_b$. The dotted horizontal lines mark the bounds obtained (i) for separable states, (ii) for states outperforming the standard teleportation scheme or (iii) for states violating the CHSH-Bell inequality.
}\label{figteleportation}
\end{figure}

\textbf{Analogy to isotropic states:} The process of the two photons generated by a para-positronium decay can be identified with a source generating an isotropic state (rotation invariant) with a weight $p= \mathcal{V}(k_i,\tilde{\Theta}_a)\mathcal{V}(k_i',\tilde{\Theta}_b)$. An isotropic state is defined as a mixture of a totally mixed state and a maximally entangled state, e.g. $\rho_{iso,p}=\frac{1-p}{4}\mathbbm{1}_4+p\;|\psi^+\rangle\langle\psi^+|$. Differently stated, the sources produces with probability $1-p$ a totally mixed state, i.e. an unpolarised state and with probability $p$ it sends a maximally entangled Bell state. This isotropic state is known to be entangled for $p>\frac{1}{3}$ (e.g. proven by $I_3$), outperforming classical teleportation protocols for $p>\frac{2}{3}$~\cite{teleportation} and violating a CHSH-Bell inequality for $p>\frac{1}{\sqrt{2}}$~\cite{chsh}. This is visualized for different energetic photons and visibilities in Fig.~\ref{figteleportation}. Assuming two photons with equal energies the entanglement is only directly revealed if the visibility is greater than $\frac{1}{\sqrt{3}}$, which corresponds to an energy lower than $k_i=1.45\equiv741keV$. Outperforming a classical teleportation scheme is given for energies lower than $k_i=0.63\equiv322keV$ and a formal violation of a CHSH-Bell inequality is given for energies lower than $k_i=0.56\equiv 286keV$.

\section{Experimental feasible entanglement witness for the decay of para-positronium}\label{secfeasibility}

Now we construct an experimentally feasible entanglement witness for the case of para-positronium decay. For that we assume that both photons have an energy of $511$keV and the detector has measured all four momentum vectors, hence of two incoming photons and two outgoing photons. The feasibility of such measurements can be found in Ref.~\cite{PolExp} and will be around few degrees for the improved J-PET.

The correlation function we construct by adding the two cases $(\phi_a,\phi_b)$ and $(\phi_a+\frac{\pi}{2},\phi_b+\frac{\pi}{2})$, where we find the opposite behaviour of the scattering cross section (e.g., minimum and maximum). The two other mutually unbiased setups we obtain by unitarily rotating the initial state from the linear polarised basis to the $45^\circ$ basis and circular basis, respectively. Then we have to maximize over all local unitaries. This does lead, however, to upper bounds on the MUB witness~(\ref{MUBwitnessFormula}) that are for separable states greater than for the maximally entangled state, a truly unphysical result. The reason being that any state, separable or entangled, of two photons has to obey also the Bose symmetry. Thus the correlation function has for both cases be extended by changing $(\theta_a,\phi_a)\leftrightarrow(\theta_b,\phi_b)$ and by commuting the initial state $\rho_{ab}$ to $\rho_{ba}$.

For the two $511keV$ photons in the $|\psi^+\rangle_{\textrm{lin}}$ state we obtain the MUB-witness for the normalized part to be $\max I_3=\frac{1}{2}(3+3\mathcal{V}(1,81,67^\circ)\mathcal{V}(1,180^\circ-81,67^\circ))=2.21789$. The maximum is obtained for the following triple of rotations (that are mutually exclusive):
\beq
\left\{\mathbbm{1}_2\otimes\mathbbm{1}_2,\;\frac{1}{\sqrt{2}}\left(\begin{array}{cc} 1&-1\\1&1\end{array}\right)\otimes \frac{1}{\sqrt{2}}\left(\begin{array}{cc} 1&-1\\1&1\end{array}\right),\right.\nonumber\\
\left.\frac{1}{2}\left(\begin{array}{cc} 1+i&1-i\\1-i&1+i\end{array}\right)\otimes \frac{1}{2}\left(\begin{array}{cc} 1+i&1-i\\1-i&1+i\end{array}\right)^*\right\}
\eeq
Any separable state obeying the Bose symmetry gives maximally $\max I_3=\frac{1}{2}(3+\mathcal{V}(1,81,67^\circ)\mathcal{V}(1,180-81,67^\circ))=1.7393$. Thus, for separabel states the assumption of rotational invariance of the state, i.e. the result given in Eq.~(\ref{IsymSEP}), is the optimal one.

Based on a similar idea one can construct entanglement witnesses based on symmetric informationally complete positive operator-valued measures (SIC-POVMs) or special generalized measures. They and the SIC-witnesses are in detailed explained in the Appendix~\ref{appendixSIC}. In the case of exploiting a full set of MUBs and SICs no difference is observed, which is related to the fact that both are $2$-designs~\cite{kdesignHiesmayr}. However, if less MUBs or less SICs are used, we observe differences. For praxis all those can be exploited in proving the nature of the quantum state.

\section{Non-trivial examples from the decay of ortho-positronium}\label{Secnontrivial}

\begin{figure}
\includegraphics[width=0.5\textwidth]{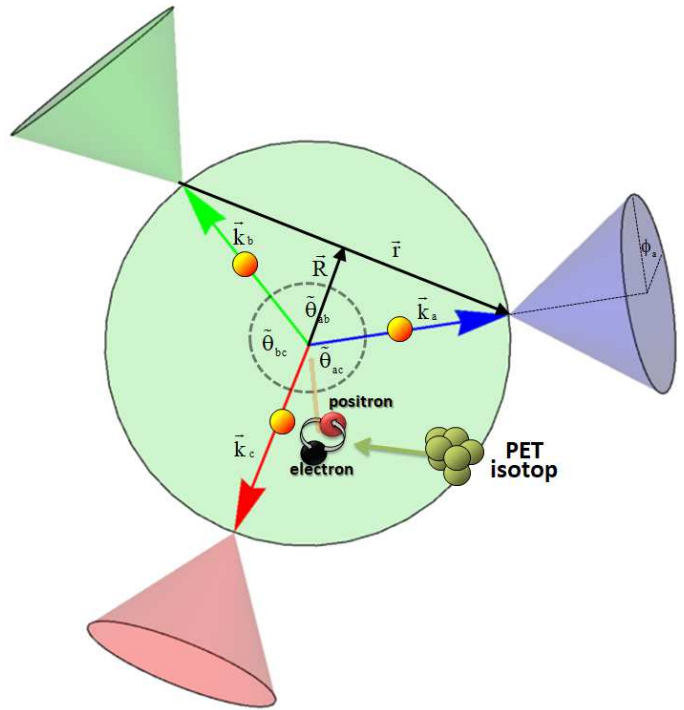}
\caption{(Color online) This picture sketches the decay of ortho-positronium into three photons as is the case when a patient is examined by a PET-radio-active substrate emitting positrons forming the positronium. Here the case is shown when all three photons have the same energy and consequently an angle of $\tilde{\theta}_{ij}=\frac{2\pi}{3}$ with $i,j$ labelling the three photons $a,b,c$. All three photons are assumed to undergo a Compton scattering in the detector under a certain scattering angle $\tilde{\Theta}_i$. The relative vector $\vec{r}$ and the $\vec{R}$ denote the center-of-mass coordinates for two photons.
}\label{figortho}
\end{figure}

In the examples above we considered a back-to-back symmetry and a source producing maximal entanglement. The product of the energy and scattering angle dependent visibilities were the main player allowing for distinguishing between separability and entanglement. Now, we consider a source producing entangled states with a non-trivial symmetry, the decay of positronium atoms into three photons sketched in Fig.~\ref{figortho}. Here, we have no selection rule at work if one considers any two photons out of three (by ignoring one photon), this bipartite state does not need to be in a particular parity eigenstate. Details are given in the Appendix~\ref{appendix1}. The entanglement properties of the three-photon decay of ortho-positronium are involved and discussed in Ref.~\cite{PositroniumHiesmayr}. In particular the three-photon states are depending on the angles between the three momentum vectors. Formula~(\ref{generalmultipartiteresult}) in Ref.~\cite{PositroniumHiesmayr} gives us the recipe how to compute the cross section given a three-photon state resulting from an ortho-positronium decay.

Let us consider the most symmetric case, i.e. all three angles between the three photons $a,b,c$ are equal $\theta_{ab}=\theta_{ac}=\theta_{bc}=\frac{2\pi}{3}$. In this case the energies have to be equal, i.e. $k_i=\frac{2}{3}$ since the total sum has to be two times the rest energy of an electron. Each single Compton event would have an optimal scattering angles of $\tilde{\Theta}=85^\circ$ for which the single visibility is maximal ($\mathcal{V}_{\textrm{max}}=0.8$). Ortho-positronium can have three different spin states $-1,0,1$. For each spin-eigenstate the entanglement content equals~\cite{PositroniumHiesmayr}, however, they differ by overall phases, in particular the role of $H$ and $V$ is changed by changing from the spin zero state to those of $\pm 1$. If no defined spin state is prepared, we expect an equal mixing of those three spin-eigenstates and in general $p$ will quantify the mixing. The reduced state for any two photons out of the three for a given spin-mixing $p$ computes to (the reference coordinate system is to be fixed along the $a$ photon)
\beq
\rho_p&=&\frac{1}{6}\;|\psi^+\rangle_{\textrm{lin}}\langle\psi^+|_{\textrm{lin}}+p_+ |\tilde\psi_+\rangle\langle\tilde\psi_+|+p_- |\tilde\psi_-\rangle\langle\tilde\psi_-|\;,\nonumber\\
\eeq
with $p_\pm=\frac{1}{12}(5\pm\sqrt{5^2+8^2p(p-1)})$ and
\beq
|\tilde\psi_\pm\rangle &=&\tiny{\frac{1}{\sqrt{1+(\frac{1}{3}(8(p-\frac{1}{2})\pm\sqrt{5^2+8^2 p(p-1)}))^2}}}\nonumber\\
\lefteqn{\left\{\frac{1}{3}(8(p-\frac{1}{2})\pm\sqrt{5^2+8^2 p(p-1)})|HH\rangle+|VV\rangle\right\}\;.}\nonumber\\
\eeq
This state $\rho_p$ has for all values of $p$ the same concurrence $\mathcal{C}=\frac{1}{3}$, i.e. equal amount of entanglement. Thus our MUB-witness should result in the same value for all weights $p$. For instance if we choose $p=\frac{1}{2}$ the state becomes a Bell-diagonal state, namely one that can be decomposed in three Bell states (i.e. for this choice the analytical computation simplifies).

To compute the MUB-witness we have to compute the (normalized) correlation function and probabilities, respectively. This has to be done by adapting the Kraus-type operators to the new geometry, namely the back-to-back geometry has to be changed to the an angle of $\frac{2\pi}{3}$, however, we can also transform to the center of mass system of the two photons (see Appendix~\ref{appendix1} for details).

Optimizing the MUB-witness for any weight $p$ one finds $I_3(\rho_p)=\frac{1}{2}(3+2 \mathcal{V}(85^\circ,\frac{2}{3})^2)=2.14$, whereas the optimization over all separable states, can be deduced from  Eq.~(\ref{IsymSEP}) resulting in $\max I_3(\rho_{\textrm{SEP}})=\frac{1}{2}(3+\mathcal{V}(85^\circ,\frac{2}{3})^2)= 1.82$ (in agreement with a numerical optimization directly from the respective Kraus operators). The visibility is increased in general, because the photon's energy is less, however, the state is less entangled, thus the witness gets in total closer to the bound. The MUB-witness optimizes for $\tilde\Theta_a=\tilde\Theta_b=85^\circ$ as we would expect from our theoretical considerations and differs in a factor of $2$ from an optimal scenario resulting from separable states for all weights $p$.

\section{Summary and Outlook}

In the seminal 1928-paper~\cite{KleinNishina} Klein and Nishina applied the new relativistic theory by Dirac to the scattering process of a photon on free electrons. By that they described also the high energetic limit correctly. We have reformulated the Klein-Nishina formula in a quantum-information theoretic way. In particular, we have singled-out the term in front of the interference term as the interference-contrast, the \textit{a priory visibility} $\mathcal{V}(k_i,\tilde{\Theta})$ of the process. This energy and scattering angle dependent expression quantifies the ability and efficiency to deduce the polarisation of the incoming photon via the outgoing Compton scattered photon. For instance, for the $511keV$ photons generated by the decay of para-positronium the maximal visibility is $0.69$ obtained for a scattering angle of $81,67^\circ$. Small angles or angles around $180^\circ$ do not allow to obtain information about the polarisation via Compton scattering processes independent of the photon energy.

On the other hand, the \textit{envelope} or \textit{shaping} function that multiplies the normalized probability part quantifies the statistics of the scattering angle in dependence of the energy. For low energetic photons small and big scattering angles are favoured equally, the higher the energy forward scattering angles become favoured over backward scattered angles. For optimizing a polarisation measurement we have a tradeoff between statistics and visibility for a given photon energy.

We have further re-written the Klein-Nishima formula via Kraus operators, namely a scattering dependent one and a polarisation dependent one. This illustrates that the Compton scattering process is an imperfect polarisation measurement apparatus. Moreover, it allows a straightforward generalization to multi-photon scattering processes of multi-photon states as it is e.g. the case in a typical PET scan in a hospital. The theory predicts that the state of a positronium is in a maximal entangled Bell state or a genuine multipartite entangled state~\cite{PositroniumHiesmayr} in the case of a two-photon or three-photon event, respectively. This work investigated how this entanglement can be experimentally proven in the case one considers a pair of two incoming and outgoing photons. The developed entanglement witness, based on mutually unbiased basis (MUB) choices, is a function of the visibilities of both scattering processes and cannot exceed a certain value for any separable state. We have shown in detail for which energies and scattering angles Compton scattering processes can be utilized to distinguish a source of separable states from a source of entangled states.

More generally, a set of non-complete MUBs or SICs, specific positive operator-valued measures, can be used to distinguish experimentally between scenarios of a source of separable states or a particular entangled state (computed in the Appendix~\ref{appendixSIC}). In short, it exploits that the choice of reference system is arbitrary on an event basis.

The quantum-information theoretic framework allows also a direct comparison to typical efficiencies of quantum protocols such as the teleportation scheme or violation of the Bell-CHSH inequality in dependence of the entanglement content (see Fig.~\ref{figteleportation}). Last but not least we have shown, how the Kraus operators allow to adapt straightforwardly to any given geometry of the photons, e.g. as in the case of the decay of ortho-positronium where the angle between the photons can have in general take any value in the interval $\{0^\circ,180^\circ\}$.

The J-PET tomograph~\cite{JPETT1,JPET1,JPETX,JPETT4,JPETT5,JPETT6}, a novel tomograph based on cutting edge technology, is able to detect the incoming and outgoing momentum vectors of high energetic photons via Compton scattering processes. Thus it has the potential to apply the proposed entanglement witnesses and by that to observe experimentally the predicted entanglement in positronium decays. Moreover, it has been shown that the positronium decay is sensitive to the surrounding, which was used to witness cancerous versus healthy human tissues taken from operated patients~\cite{Cancer1,Cancer2}. Further differences of the lifetimes with respect to the spin states of positronium are recorded when exposed to light or not~\cite{Cancer3}. Thus, positronium is sensitive to its environment, therefor one may expect that the entanglement features are also sensitive to biological effects and herewith opening a fully novel door for observing entanglement features in living organisms.

\vspace{0.5cm}

\section*{Acknowledgements}
B.C.H. acknowledges gratefully the Austrian Science Fund (FWF-P26783) and P.M. acknowledges support by the Foundation
for Polish Science through the TEAM/2017-4/39 project, and the National Science Centre of Poland through grant no. 2016/21/B/ST2/01222.

\appendix

\section{Two-photon wave functions}\label{appendix1}

A single photon state can be described by the (tensor) product of the spatial and internal degrees of freedom~\cite{HiesmayrSinglePhoton}, i.e. in our formalism photon $a$ is described by
\beq
e^{-i \omega_a t} e^{i \bf{k}_a \bf{r}_a }\otimes \boldsymbol{\varepsilon}({\bf \hat{k}}_a,\lambda_a)\;.
\eeq
Any multipartite photon state can be obtained by tensoring the one-particle states. Let us consider two photons $a$ and $b$ and their possible states (suppressing the tensor product between the spatial and inner degrees of freedom but not those between the individual photons with respect to the inner degrees of freedom)
\beq
&&e^{-i \omega_a t} e^{i  \bf{k}_a\bf{r}_a}  e^{-i \omega_b t} e^{i \bf{k}_b\bf{r}_b }\; {\bf \varepsilon}({\bf \hat{k}}_a,\lambda_a)\otimes {\boldsymbol\varepsilon}({\bf \hat{k}}_b,\lambda_b)\\
&=& e^{-i (\omega_a+\omega_b) t} e^{i  (\bf{k}_a+\bf{k}_b)\bf{R}} e^{i  (\bf{k}_a-\bf{k}_b)\frac{{\bf r}}{2}}\; {\boldsymbol\varepsilon}({\bf \hat{k}}_a,\lambda_a)\otimes {\boldsymbol\varepsilon}({\bf \hat{k}}_b,\lambda_b)\;,\nonumber
\eeq
where we have defined the center-of-mass coordinates by $\bf{R}=\frac{{\bf r}_a+{\bf r}_b}{2},\;\bf{r}={\bf r}_a-{\bf r}_b$ (see also Fig.~\ref{figortho}).

\begin{widetext}
Since photons are bosons, we have to symmetrize the wave function (${\bf k}:=(\bf{k}_a-{\bf k}_b)/2$)
\beq
&&e^{-i (\omega_a+\omega_b) t} e^{i  (\bf{k}_a+\bf{k}_b)\bf{R}}\cdot
\left(e^{i  {\bf k}{\bf r}}\; {\boldsymbol\varepsilon}({\bf \hat{k}}_a,\lambda_a)\otimes {\boldsymbol\varepsilon}({\bf \hat{k}}_b,\lambda_b)+e^{-i  {\bf k}{\bf r}}\; {\boldsymbol\varepsilon}({\bf \hat{k}}_a,\lambda_b)\otimes {\boldsymbol\varepsilon}({\bf \hat{k}}_b,\lambda_a)\right)\;=\nonumber\\
&&\frac{1}{2}\;e^{-i (\omega_a+\omega_b) t} e^{i  (\bf{k}_a+\bf{k}_b)\bf{R}}\cdot\\&&
\left[\left(e^{i  {\bf k}{\bf r}}+e^{-i  {\bf k}{\bf r}}\right)\left\lbrace{\boldsymbol\varepsilon}({\bf \hat{k}}_a,\lambda_a)\otimes {\boldsymbol\varepsilon}({\bf \hat{k}}_b,\lambda_b)+ {\boldsymbol\varepsilon}({\bf \hat{k}}_a,\lambda_b)\otimes {\boldsymbol\varepsilon}({\bf \hat{k}}_b,\lambda_a)\right\rbrace\right.\nonumber\\
&&\left.
+
\left(e^{i  {\bf k}{\bf r}}-e^{-i  {\bf k}{\bf r}}\right)\left\lbrace{\boldsymbol\varepsilon}({\bf \hat{k}}_a,\lambda_a)\otimes {\boldsymbol\varepsilon}({\bf \hat{k}}_b,\lambda_b)- {\boldsymbol\varepsilon}({\bf \hat{k}}_a,\lambda_b)\otimes {\boldsymbol\varepsilon}({\bf \hat{k}}_b,\lambda_a)\right\rbrace\right]\nonumber\;.
\eeq
Now, any polarisation vector can be decomposed by two orthonormal vectors, i.e. ${\boldsymbol\varepsilon}({\bf \hat{l}},\lambda_b)= c_1({\bf \hat{l}})\, {\boldsymbol\varepsilon}({\bf \hat{l}},\lambda_a)+c_2({\bf \hat{l}})\, {\boldsymbol\varepsilon}({\bf \hat{l}},-\lambda_a)$ with $c_i$ being complex numbers with $|c_1|^2+|c_2|^2=1$. Thus we can re-write the polarisation dependent part by
\beq
\{c_1({\bf \hat{k}}_b)\pm c_1({\bf \hat{k}}_a)\}{\boldsymbol\varepsilon}({\bf \hat{k}}_a,\lambda_a)\otimes {\boldsymbol\varepsilon}({\bf \hat{k}}_b,\lambda_a)+ c_2({\bf \hat{k}}_b){\boldsymbol\varepsilon}({\bf \hat{k}}_a,\lambda_a)\otimes {\boldsymbol\varepsilon}({\bf \hat{k}}_b,-\lambda_a)\pm c_2({\bf \hat{k}}_a){\boldsymbol\varepsilon}({\bf \hat{k}}_a,-\lambda_a)\otimes {\boldsymbol\varepsilon}({\bf \hat{k}}_b,\lambda_a)\;.
\eeq
\end{widetext}
If we move to the center of mass system, then the momenta of both photons have opposite sign and same amount which we denote by the unit vector ${\bf \hat{l}}$. Now a vector orthonormal to the vector ${\boldsymbol\varepsilon}({\bf \hat{l}},\lambda_b)$ is up to total phase of $-1$ equal to ${\boldsymbol\varepsilon}({\bf \hat{l}},-\lambda_b)= -c_2({\bf \hat{l}})^*\, {\boldsymbol\varepsilon}({\bf \hat{l}},\lambda_a)+c_1({\bf \hat{l}})^*\, {\boldsymbol\varepsilon}({\bf \hat{l}},-\lambda_a)$, which transforms with the general property $\varepsilon({\bf \hat{l}},\lambda)=-\varepsilon(-{\bf \hat{l}},-\lambda)$ to a state moving in the opposite direction ${\boldsymbol\varepsilon}(-{\bf \hat{l}},\lambda_b)= c_1({\bf \hat{l}})^*\, {\boldsymbol\varepsilon}(-{\bf \hat{l}},\lambda_a)-c_2({\bf \hat{l}})^*\, {\boldsymbol\varepsilon}(-{\bf \hat{l}},-\lambda_a)$, which by setting ${\bf \hat{l}}=-{\bf \hat{h}}$ turns to ${\boldsymbol\varepsilon}({\bf \hat{h}},\lambda_b)= c_1(-{\bf \hat{h}})^*\, {\boldsymbol\varepsilon}({\bf \hat{h}},\lambda_a)-c_2(-{\bf \hat{h}})^*\, {\boldsymbol\varepsilon}({\bf \hat{h}},-\lambda_a)$, which expresses that the orthogonal state moves in opposite direction is given by same circular polarized state, the weights undergo a complex conjugation and a relative phase change. Consequently, our polarisation dependent part in the center of mass system turns into
\beq
|\lambda \lambda\rangle\cdot\left\lbrace\begin{array}{c} 2 Re\{c_1\}\\ 2 Im\{c_1\}\end{array}\right.-Re\{c_2\} |\psi^\mp\rangle_{\textrm{circ}}+i Im\{c_2\} |\psi^\pm\rangle_{\textrm{circ}}\;,\nonumber\\
\eeq
where we used the abbreviation $|\lambda\lambda\rangle\equiv {\boldsymbol\varepsilon}({\bf \hat{k}},\lambda)\otimes {\boldsymbol\varepsilon}(-{\bf \hat{k}},\lambda)$ and the $|\psi^\pm\rangle$ denote the Bell states. For the part proportional to ${\bf R}$ we have same sign of the momenta of the two photons, consequently this polarisation dependent part becomes
\beq
|\lambda \lambda\rangle\cdot\left\lbrace\begin{array}{c} 2 c_1\\ 0\end{array}\right.+c_2 |\psi^\pm\rangle_{\textrm{circ}}\;.
\eeq

Our next goal is to assure that $\psi_{ab}$ is an eigenstate of the parity operator $\bf{\mathcal{P}}$. The operator does not affect the spatial part since the product $\bf{\mathcal{P}}\bf{k}\bf{r}=(-\bf{k})(-\bf{r})=\bf{k}\bf{r}$ does not change the sign. A parity operation onto the circular polarised states leads to ${\bf\mathcal{P}}\,|\lambda\rangle=-|-\lambda\rangle$. This can be seen from $\bf{\mathcal{P}}\,{\boldsymbol\varepsilon}({\bf \hat{k}},\lambda)={\boldsymbol\varepsilon}({-\bf \hat{k}},\lambda)=-{\boldsymbol\varepsilon}({\bf \hat{k}},-\lambda)$. Thus we find that the contribution $|\lambda\lambda\rangle$ is not an eigenstate of the parity operator, hence $c_1$ has to be zero, which means that the polarisation of the photon $b$ has to be orthogonal to the one of the photon $a$. Therefore, $|c_2|^2=1$. If we choose $c_2=\pm 1$ the two terms in the sum do no longer separately obey the Bose symmetry, which is only the case if we choose $c_2=\pm i$. Thus the wave-functions obeying the Bose symmetry and corresponding to an eigenvalue of the parity operator $\bf{\mathcal{P}}$ are maximal entangled states (up to overall phases and normalisation)
\beq
\psi_{ab}^{\mathcal{P}=+1}&=& \frac{1}{2} e^{-i (\omega_a+\omega_b) t}\cdot e^{i  (\bf{k}_a+\bf{k}_b)\bf{R}}\cdot\\
&&\left\lbrace e^{i  \bf{k}\bf{r}}+e^{-i  \bf{k}\bf{r}}\right\rbrace\cdot \left\{|\lambda, -\lambda\rangle+|-\lambda, \lambda\rangle\right\}\nonumber\\
&\equiv&\frac{1}{\sqrt{2}} e^{-i  (\omega_a+\omega_b) t} \cdot e^{i  (\bf{k}_a+\bf{k}_b)\bf{R}}\cdot\nonumber\\
&&\left\lbrace e^{i  \bf{k}\bf{r}}+e^{-i  \bf{k}\bf{r}}\right\rbrace\otimes |\psi^+\rangle_{\textrm{circ}}\nonumber\\
\psi_{ab}^{\mathcal{P}=-1}&=& \frac{1}{2} e^{-i  (\omega_a+\omega_b) t} e^{i  (\bf{k}_a+\bf{k}_b)\bf{R}}\cdot\\
&&\left\lbrace e^{i  \bf{k}\bf{r}}-e^{-i  \bf{k}\bf{r}}\right\rbrace\cdot\left\lbrace |\lambda, -\lambda\rangle-|-\lambda, \lambda\rangle\right\rbrace\nonumber\\
&\equiv&\frac{1}{\sqrt{2}} e^{-i  (\omega_a+\omega_b) t}\cdot e^{i  (\bf{k}_a+\bf{k}_b)\bf{R}}\cdot\nonumber\\
&& \left\lbrace e^{i  \bf{k}\bf{r}}-e^{-i  \bf{k}\bf{r}}\right\rbrace\otimes |\psi^-\rangle_{\textrm{circ}}\nonumber
\eeq
In a spin-singlet state the ground state of the electron-positron bound system has parity $\mathcal{P}=-1$, since parity is conserved in electromagnetic interactions the final state of the photons have to have $\mathcal{P}=-1$. We have defined the polarisation with respect to the propagating direction $\pm {\bf \hat{k}}$, respectively. Returning to one reference system we obtain the Bell state $|\psi^-\rangle_{\textrm{circ}}\longrightarrow |\phi^-\rangle_{\textrm{circ}}$, which corresponds to $|\psi^+\rangle_{\textrm{lin}}$ in the linear polarised basis. The result for $|\psi^+\rangle_{\textrm{lin}}$ is also reported in e.g. Ref.~\cite{Harpen,Snyder}.

\section{An entanglement witness based on symmetric informationally complete generalized quantum measurements}\label{appendixSIC}

Mutually unbiased bases obey the relation  $|\langle i_l^{(k)}| i_{l'}^{(k')}\rangle|^2 \;=\; \delta_{l,l'}\delta_{k,k'}+(1-\delta_{k,k'})\frac{1}{d}$. A transformation from one MUB to another forms a complex unitary Hadamard matrix $\mathcal{H}$ with unipolar entries $|\mathcal{H}_{ij}|=\frac{1}{d}$. The classification of all Hadamard matrices is only known up to $d=5$~\cite{Haagerup}. Surprisingly, a complete set of MUBs, namely $d+1$ bases are only known in dimension of prime numbers or power-prime numbers. For instance, in the first non-trivial dimension $d=6$ we would expect $7$ MUBs, however, so far only $3$ MUBs have been found and numerical considerations suggest that there exists no forth one, however, though a long standing problem in literature, no analytical proof exists.

\begin{figure}
\includegraphics[width=0.35\textwidth]{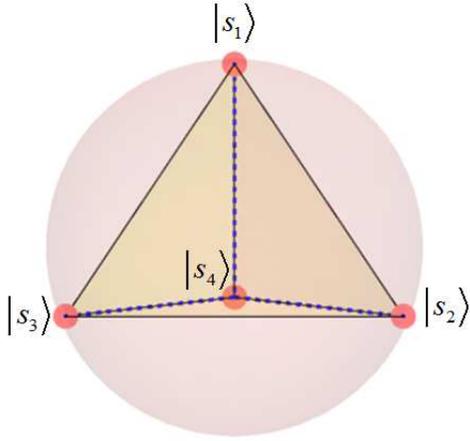}
\caption{(Color online) This picture sketches the Bloch's sphere or ball, where each point can be associated with a qubit-state. In particular, any pure state can be identified with a point on the surface, whereas mixed states are within the ball. Any unitary operation ``rotates'' a pure state to another point on the surface of the ball. If four points at the surface form a regular tetrahedron, then they are symmetric informationally complete.
}\label{figureSIC}
\end{figure}

For $d$-dimensional systems symmetric informationally complete generalized quantum measurements or symmetric informationally complete positive operator-valued measures (SIC-POVMs) or shorter SICs, are sets of POVMs which obey
\beq
|\langle s_l| s_{l'}\rangle|^2 &=& \frac{1}{1+d}\qquad\textrm{with}\quad l\not=l'\;,
\eeq
e.g. weighted projections. They are conjectured to exist in all dimensions~\cite{Grassl1,Grassl2}. For qubits, $d=2$, such a set of vectors forms a tetrahedron in the Bloch's sphere, visualized in Fig.~\ref{figureSIC}. Starting with an arbitrary seed state $s_1$, we find the remaining three SIC states by applying, e.g., the following three unitary matrices
\beq\label{optmatrices}
\{\frac{1}{\sqrt{3}}\left(\begin{array}{cc}1&\sqrt{2}\\\sqrt{2}&-1\end{array}\right),
\left(\begin{array}{cc}1&\sqrt{2}\\-(-1)^\frac{1}{3}\sqrt{2}&(-1)^\frac{1}{3}\end{array}\right),\nonumber\\
\left(\begin{array}{cc}1&\sqrt{2}\\(-1)^\frac{2}{3}\sqrt{2}&-(-1)^\frac{2}{3}1\end{array}\right)\}\;.
\eeq

Via these SICs we can construct a similar protocol~\cite{BaeSIC,kdesignHiesmayr} as the one with MUBs discussed in the main text. Namely that for any separable state the following inequalities hold
\beq
L_{\tilde{m}}\;\leq\;\frac{2}{3} \sum_{i=1}^{\tilde{m}}Tr\{|s_i\rangle\langle s_i|\otimes\sigma_z|s_i\rangle\langle s_i|\sigma_z\;\rho\}\;\leq\;U_{\tilde{m}}\;.
\eeq
Here the third Pauli-matrix $\sigma_z$ is need such that $|\psi^+\rangle$ is the optimum (and not $|\psi^-\rangle$ or $|\phi^+\rangle$) for the above choice of matrices~(\ref{optmatrices}). The lower ($L$) and upper bounds ($U$) are depending on the number of SIC-vectors $\tilde{m}$, in detail one finds for increasing $\tilde{m}$ for the lower bound $L=\{0,0,\frac{2}{5},1\}$  and  for the upper bound $U=\{0,(\frac{1+\sqrt{3}}{3})^2, 2,2\}$. Since a complete $d^2$ set of SICs is related by a $k$-design to a complete set of $d+1$ MUBs, the above inequality is identical to the one for MUBs, Eq.~(\ref{MUBwitnessFormula}). Note also for multipartite system information complete quantum measurements may be of interest~\cite{Karol}.

\begin{widetext}
\begin{center}
\begin{table}[h!]
\caption{This table summarizes the values obtained for different numbers of MUBs or SICs for the upper and lower bounds for low energetic and high energetic photons. Violations are highlighted by a bold font.}\label{tableMUBSIC}
\setlength{\arrayrulewidth}{0.8pt}
\setlength{\tabcolsep}{15pt}
\renewcommand{\arraystretch}{1.5}
\begin{tabular}{||c|c|c|c|c|c||}
\hhline{|t:======:t|}
 & & \multicolumn{2}{c|}{MUB-witness}
& \multicolumn{2}{c||}{SIC-witness}\\
\hhline{||cccc||}
 energy & number & \multicolumn{2}{c|}{(\textit{lower}/\textit{upper})} & \multicolumn{2}{c||}{(\textit{lower}/\textit{upper})}\\
 \hhline{||-|-|-|-|-|-||}
 & ($m$,$\tilde{m})$ & SEP (opt.)& ENT ($|\psi^+\rangle$)& SEP(opt.) &ENT ($|\psi^+\rangle$)\\
  \hhline{||=|=|=|=|=|=||}
$k_i=0.0001$ & (3,4) & 1/2&{\bf 0}/{\bf 3}   &1/2 &{\bf 0}/{\bf 3}\\
           & (2,3) & 0.5/1.5&{\bf 0}/{\bf 2}  & $\frac{2}{5}$/2 &{\bf 0}/{\bf 2.25}\\
           & (1,2) & 0/1&0/1 &0/$(\frac{1+\sqrt{3}}{2})^2$ &0/1.5\\
  \hhline{||=|=|=|=|=|=||}
$k_i=1$ & (3,4) & $1.26$/$1.74$&\bf{0.78}/{\bf 2.22}   &1.26/1.74 &{\bf 0.78}/{\bf 2.22}\\
           & (2,3) &$0.76$/$1.24$  &{\bf 0.52}/{\bf 1.48} & 0.66/1.70& {\bf 0.59}/ 1.66\\
           & (1,2) & $0.26$/$0.74$&  $0.26/0.74$          &0.26/1.47&0.39/1.11\\
\hhline{|b:======:b|}
\end{tabular}
\end{table}
\end{center}
\end{widetext}

In the Table~\ref{tableMUBSIC} we have summarized the result for $511keV$ photons and low energetic ones (visibility close to one) optimized over all remaining parameters for an optimal separable state and an initial maximally entangled state including local unitary matrices~\cite{compositeparamerization} (numbers in bold show a violation).

Obviously, one cannot witness entanglement if there are less than two MUBs or three SICs employed. For the MUB-witness the lower and upper bounds decrease symmetrically with the visibility. This is not the case for the SIC-witness since three SICs are not enough to violate the upper bound, but the lower bound is violated.

\end{document}